\documentclass{aa} 
\pdfoutput=1
\usepackage{graphicx}
\usepackage{txfonts}
\usepackage{slashbox}
\usepackage{natbib}
\begin{document}

\title{Structure and evolution of the photospheric magnetic field in 2010 - 2017: comparison of SOLIS/VSM vector field and $B_{LOS}$ potential field.}

\author{Ilpo I. Virtanen\inst{1}
\and
Alexei A. Pevtsov\inst{1,}\inst{2,}\inst{3}
\and
Kalevi Mursula\inst{1}
}

\institute{ReSoLVE Centre of Excellence, Astronomy and Space Physics research unit, University of Oulu, POB 3000, FIN-90014, Oulu, Finland\\
\email{ilpo.virtanen@oulu.fi, kalevi.mursula@oulu.fi}
\and
National Solar Observatory, Boulder, CO 80303, USA\\
\email{apevtsov@nso.edu}
\and
Pulkovo Astronomical Observatory, Russian Academy of Sciences, Saint Petersburg, 196140, Russian Federation\\
 }

\date{Received ; accepted}
  \abstract
   {
The line-of-sight (LOS) component of the large-scale photospheric magnetic field has been observed since the 1950s, but the daily full-disk observations of the full vector  magnetic field started only in 2010 using the SOLIS Vector Stokes Magnetograph  (VSM)  and the SDO helioseismic and magnetic imager (HMI). 
Traditionally, potential field extrapolations are based on the assumption that the magnetic field in the photosphere is approximately radial. The validity of this assumption has not been tested yet.
   }
   {
We investigate here the structure and evolution of the three components of the solar large-scale magnetic field in 2010 - 2017, covering the ascending to mid-declining phase of solar cycle 24, using SOLIS/VSM vector synoptic maps of the photospheric magnetic field. 
   }
   {
We compare the observed VSM vector magnetic field to the potential vector field derived using the VSM LOS  magnetic field observations as an input. 
The new vector field data allow us to derive the meridional inclination and the azimuth angle of the magnetic field and to investigate their solar cycle evolution and latitudinal profile of these quantities.
   }
   {
SOLIS/VSM vector data show that the photospheric magnetic field is in general fairly non-radial.
In the meridional plane the field is inclined toward the equator, reflecting the dipolar structure of the solar magnetic field.
Rotationally averaged meridional inclination does not have significant solar cycle variation.
While the vector radial component $B_r$ and the potential radial component $B_r^{PFSS}$ are fairly similar, the meridional and zonal components do not agree very well.
We find that SOLIS/VSM vector observations are noisy at high latitudes and suffer from the vantage point effect more than LOS observations. 
This is due to different noise properties in the LOS and transverse components of the magnetic field, which needs to be addressed in future studies.
   }
        {}

\titlerunning{Vector magnetic field}
\authorrunning{Virtanen, Pevtsov and Mursula}

\keywords{Sun: magnetic fields, Sun: activity, Sun: photosphere}

\maketitle

\section{Introduction} \label{sec:intro}

Maps of the photospheric magnetic field (or simply magnetograms) based on the Zeeman effect have been observed since the mid-1950s (e.g., \cite{Babcock_1953,Severny1964}).  The first magnetographs only measured  the circular polarization, which allowed the derivation of the line-of-sight (LOS) component of the magnetic field. 
Observations of the full vector magnetic field in the photosphere and chromosphere started in the early 1960s, but were limited to small areas, usually active regions. 
Full-disk magnetograms exist since the late 1960s  \citep{Howard1976,Livingston.etal1976}, typically measuring the LOS component only. 
Regular full-disk observations of the whole magnetic vector field began in 2003 with the Vector Stokes Magnetograph 
(VSM) on the Synoptic Optical Long-Term Investigations of the Sun (SOLIS) platform \citep{Keller.etal2003,Balasubramaniam.Pevtsov2011}. 
Since 2010, full-disk photospheric vector magnetograms are also routinely made  by the Helioseismic and Magnetic Imager \citep[HMI,][]{Scherrer.etal2012} on board the Solar Dynamics Observatory (SDO) satellite.
Based on the full-disk observations by these two instruments, rotational synoptic maps of the three components of the magnetic field vector since 2010 were recently constructed \citep{Gosain.etal2013,Hughes.etal2016}.
We note  that SOLIS/VSM observations in 2003 -- 2009 were too sparse to construct synoptic maps.
The full vector magnetic field is needed, for example  to better understand the structure of the photospheric magnetic field and as an improved boundary condition for coronal and heliospheric models. 

Before  routine vector field observations, the LOS component of the photospheric magnetic field has been used in modeling and other studies for several decades.
When constructing synoptic maps based on the LOS component, the assumption is made that the photospheric magnetic field is radial.
Under this assumption the radial component of the magnetic field can be solved by simply dividing the LOS field by the cosine of the angle from the center of the full-disk observation, the so-called heliocentric angle \citep{Duvall_1979, Svalgaard_1978}. 
This field is commonly called the pseudo-radial field, $B_r^{ps}$.
In fact, most solar observatories publish synoptic maps of $B_r^{ps}$, not $B_{LOS}$  \citep{Riley_2014, Virtanen_2016}.

The horizontal components of the magnetic field vector could also be derived from the LOS magnetograms  under the restrictive assumption that most changes that are seen in a selected area of the Sun are only due to the projection of the same vector field, not due to its temporal evolution.
The average zonal (east - west) component of the field, $B_\phi,$ can be derived from the B$_{LOS}$ by comparing observations of the same region over several days (at different vantage points in longitude) \citep{Howard_1974,Svalgaard_1978,Ulrich_2006}.
This method allows the zonal field averaged over several days to be solved , since the solar longitude changes only about 14$^\circ$ per day. 
Based on this approach, it was found that the orientation of the zonal field in the active regions follows the Hale polarity law of the ongoing solar cycle, being positive, i.e., westward oriented (negative, eastward oriented) in the north, and negative (positive) in the south for even (odd) cycles.
Moreover, observations of photospheric magnetic field outside sunspots have shown that, for example, a negative zonal field in the north appears at high latitudes typically  during the early declining phase of even solar cycles when the active region zonal field at the north is systematically positive. 
This depicts that this zonal cycle starts at high latitudes already after the sunspot maximum of the previous solar cycle, and  takes about 1.5 solar cycles until it reaches the equator at the solar minimum \citep{Ulrich_1993,Shrauner_1994, Ulrich_2005,Lo_2010}.
Accordingly, during the minimum of solar cycle 22 in 1996 a systematic negative large-scale $B_\phi$ in the north and positive $B_\phi$ in the south was observed  \citep{Pevtsov.Latushko2000}.
At the same time, those old-cycle active regions at low latitudes that still existed depicted positive $B_\phi$ in the north and negative in the south \citep{Ulrich_2005}.
This reflects the evolution of the zonal field where the new zonal cycle starts at high latitudes soon after the previous sunspot maximum, developing to an active region at mid-latitudes which follows the Hale law of the ongoing cycle. 

In the case of a unipolar (open) magnetic field configuration, a systematic $B_\phi$ would indicate that the magnetic field is, on average, inclined from radial in the zonal direction.
On the other hand, in the case of closed magnetic loops $B_\phi$ may have a systematic large-scale pattern, where the two ends of the loop tend to have an oppositely signed zonal inclination.
For such a closed field configuration the average radial field and the net zonal inclination are typically close to zero.
The net value of zonal inclination depends on data set used and field strength studied, with estimates varying from 0.6$^\circ$ west \citep{Shrauner_1994} to 1.9$^\circ$ east \citep{Howard_1991}.
A systematic zonal inclination of large-scale magnetic fields may result in underestimating or over-estimating the pseudo-radial field $B_r^{ps}$, depending on the angle between the line of sight and true vector magnetic field. 
However, observed zonal inclinations are fairly small, and  one could also argue that by itself, such an zonal inclination derived from the time-evolution of $B_{LOS}$ could be an artifact.
As mentioned above, the derivation of $B_\phi$ from $B_{LOS}$ is based on the assumption that the magnetic field does not change over the period of several days. 
If, however, the field does change, the derived field would appear slightly tilted eastward (westward) if the magnetic flux increases (decreases) over the period of observations.

The meridional (north - south)  field component $B_\theta$ cannot be easily estimated from B$_{LOS}$ by comparing consecutive observations, since the latitude of the vantage point changes only from $-$7.25$^\circ$ to 7.25$^\circ$ over six months.
The change over several days is too small to produce a measurable change in projection in the north-south direction.
There are some studies where $B_\theta$ has been derived using this method \citep{Wang.Zhang2010}, but it is unclear how much the variation in  $B_\theta$  is due to the varying latitude of the vantage point and how much to the evolution of the magnetic field.
Some other approaches based on LOS  observations have been introduced to estimate the possible poleward or equatorward (meridional) inclination of the field, but results  do not agree.
\cite{Ulrich_2013} corrected the polar field observations in order 
to reduce the annual vantage point effect (also called the $b_0$-angle effect) in synoptic data and concluded that high-latitude fields are inclined poleward, but the angle is rather small and confined to very high latitudes.
However, other studies found that the polar field is approximately radial \citep[see, e.g., ][and references therein]{Petrie_2015}.
We note that the  conclusions of \cite{Petrie_2015} do not take into account the \cite{Ulrich_2013}  results.
The presence of a systematic meridional inclination of the photospheric magnetic field would also affect the 
derivation of the pseudo-radial field. 
For example, in the case of equatorward inclination, $B_\theta$ would increase $B_{LOS}$ and $B_r^{ps}$ would be larger than true radial $B_r$.
In an opposite situation of poleward inclination, $B_r^{ps}$ would be smaller than the true radial $B_r$.
The effect of a possible meridional inclination to $B_r^{ps}$ would increase toward the poles. 

Polar fields are difficult to observe due to the partial invisibility of poles, but they are highly important for coronal and heliospheric models \citep{Bertello_2014, Virtanen_2016}, and  for dynamo theory \citep[see, e.g., ][and references therein]{Petrie_2015, Munoz-Jaramillo_2013}. 
Polar fields are in general rather weak, and the LOS projection of the radial field approaches  zero at high latitudes.
Vector field observations are expected to offer a solution to this problem, since vector magnetographs can also measure the transverse field, the component perpendicular to the line of sight, which may closely align with the radial polar field at high latitudes.
However, the transverse field component has a much lower signal-to-noise ratio than the LOS component. 
Therefore, it is unclear how much new information on the polar fields can be derived from vector field observations. 

Coronal and heliospheric models are the primary use of the synoptic maps of the photospheric magnetic field.
The most widely used model is the the potential field source surface  (PFSS) model \citep{Altschuler_1969,Schatten_1969,Hoeksema_1983}.
PFSS is commonly used since the more complicated models with a larger number of free parameters still cannot give a better agreement with the observed solar wind and heliospheric magnetic field than PFSS \citep{Wiegelmann_2015}.
Observations of the vector magnetic field can be used to improve the boundary conditions of the PFSS model, and to estimate how reliably the PFSS model reproduces the $B_\theta$ and $B_\phi$ components of the photospheric magnetic field, when using $B_r^{ps}$ as an input.

In this paper we employ the vector field synoptic magnetograms to investigate how the orientation of the photospheric magnetic field evolves from January 2010, the early ascending phase of solar cycle 24, to May 2017, the middle declining phase of cycle 24  (Carrington rotations 2092 -- 2190).
We study how much the photospheric magnetic field is inclined from the radial direction overall, and how much in the meridional direction. 

The paper is organized as follows.
Section 2 presents the SOLIS/VSM data and the methods used in this study. 
Section 3 shows an example of the vector magnetic field for Carrington rotation 2170 and compares it to the PFSS model. 
Section 4 presents the evolution of the large-scale vector magnetic field in 2010 - 2017 based on SOLIS/VSM vector observations and on the LOS PFSS model result. 
Section 5 discusses the inclination from the radial, meridional inclination, and zonal angle of the magnetic field and presents their long-term evolution.  
Section 6 shows the average inclination from the radial and meridional inclination of the photospheric magnetic field, and the meridional inclination for strong fields.
We discuss our results in Section 7 and give our final conclusions in Section 8.

\section{Data and methods}

We use observations of the solar vector magnetic field from SOLIS/VSM \citep{Keller.etal2003,Balasubramaniam.Pevtsov2011}.
VSM observes full line profiles of Fe I 630.15 and 630.25 nm spectral lines, with the spectral sampling of 2.3 pm and spatial sampling of 1.14'' per pixel over 2048 $\times$ 2048 pixels field of view.
To construct a full-disk magnetogram, the image of the Sun is scanned in the direction perpendicular to the spectrograph slit. 
At each scanning step (lasting about 0.6 seconds), the spectra for each pixel 
along spectrograph slit are recorded simultaneously. It takes about 20 minutes to observe one full-disk magnetogram.
The observed profiles of Stokes Q, U, V, and I parameters are inverted using the Very Fast Inversion of Stokes Vector code \citep[VFISV,][]{Borrero.etal2011} under the assumption of a standard Milne-Eddington stellar atmosphere. 
Unlike the version of VFISV code employed for HMI data reduction, the VSM inversion includes the magnetic field filling factor as an additional fit parameter. 
For a more detailed description of SOLIS/VSM inversion methods and pipeline, see \cite{Harker2017}.
During the preparation of this article for publication, the VSM inversion was modified to derive the magnetic field parameters based on the best simultaneous fit to both  Fe {\rm I} 630.15 and 630.25 nm spectral lines \citep[][ data processing level PROVER0 = 17.0615]{Harker2017}. 
New pipeline reductions started only in 2017 and the data prior to 2017 has not been re-processed using the new method.
Therefore, in order to have the best possible data coverage and homogeneous  data set,  we here employ the data from the previous version until Carrington rotation 2190 (May 2017) in which the magnetic field properties were derived based on the best fit of only one of these two Fe spectral lines (data processing level PROVER0 = 15.0511). 

The magnetic field vector is defined by its strength, inclination angle relative to the line of sight, and the azimuth angle in the plane of sky.  
The azimuth angle has 180$^\circ$ ambiguity, i.e., the transverse field has two possible directions, aligned 180$^\circ$  from each other.  
Ambiguity is resolved using the Very Fast Disambiguation Method \citep[VFDM,][]{Rudenko_2014}. 
This disambiguation method was compared with several other methods, and was found to be in a good agreement with the minimum energy method, which is perhaps the most trusted disambiguation method used, for example to derive HMI vector fields in active regions \citep{Metcalf_1994,Leka_2009}.
However, the minimum energy method is computationally very expensive, which makes it less attractive for  implementation to full-disk data. 
In the case of SOLIS/VSM, the VFDM has a similar accuracy to that of the minimum energy method, but is much faster \citep{Rudenko_2014}.

Rotational synoptic (sine-latitude -- longitude) maps of the three vector components of the photospheric magnetic field were recently constructed from daily SOLIS/VSM vector observations.
Full-disk observations are first mapped to the spherical coordinate system, where the radial component ($B_r$) is positive when pointing away from the Sun, the meridional component ($B_\theta$) is positive southward, and the zonal component ($B_\phi$) is positive westward, forming a local right-handed coordinate system $(B_r,B_\theta,B_\phi)$.\footnote{In the absence of consistent definition, past studies used the terms ``poloidal and toroidal'' and ``meridional and azimuthal'' in their reference to north-south and east-west components.
\cite{Thompson_2006}  proposed using a unified coordinate system for solar observations, but no names were suggested for the north-south and east-west components. 
Here we adopt the terminology used in meteorology where ``zonal'' means  along the latitudinal circle, and  ``meridional''  means along the longitudinal circle.}
Full-disk observations are then weighted using a $\cos^4 (\phi_{CMD})$ mask, where  $\phi_{CMD}$ is the angle in longitude from central meridian. 
This method gives a large weight to observations around the central meridian and a very small weight to observations close to the limb. 
Then the synoptic map is constructed using all the full-disk observations from 8 days before the start of Carrington rotation to 8 days after the end of the rotation. 
Additional details about the algorithm for creating the synoptic maps can be found in  \cite{Bertello_2014}.
Maps are available in two resolutions, 180 by 360 (sine-latitude -- longitude) pixels and  900 by 1800 pixels. 
In this work we use the 180 by 360 pixel maps of the vector field. 
We use here SOLIS/VSM observations since instrument sensitivity is higher than in SDO/HMI due to the larger pixel size.
This is particularly important for transverse fields, which have a higher noise than the LOS field.
Magnetograph noise levels may not be straightforward to determine since  they can vary across the field of view and depend on seeing condition (for ground-based observations). 
The effect of noise may be indirect (by affecting the shape of spectral line profiles), and difficult to separate from a real signal, for example of a quiet Sun.
Using the width of distribution of weak fields as a proxy for noise, \citet{Pietarila_2012} estimated the noise level in SOLIS/VSM $B_{LOS}$ data to be about 1G. 
SOLIS/VSM data is also corrected for stray light (see page 20 of \cite{Harker2017}). The
SDO/HMI $B_{LOS}$ noise level is about 10G \citep{Liu_2012}.
The noise level in the transverse component is expected to be about 10 times larger than  $B_{LOS}$  noise.
\cite{Thalmann_2012} found the value of 100G for HMI transverse field noise and 70G for VSM transverse field noise.
However, these values, especially the VSM value, are probably overestimated. 
The amplitude of noise in synoptic maps is reduced approximately by $\sqrt{N}$, where N is the number of  pixels in original full-resolution image (known as the  sky image) contributing to a single pixel in the synoptic map. 
For one 1$^\circ$ $\times$ 1$^\circ$ pixel of the synoptic map, there are  about 225 1.14'' full-disk sky image pixels.

Our data set spans Carrington rotations 2092--2190.
The average number of full disk observations (MAPCOUNT in fits header) used to construct a  synoptic map is 51.
The best coverage of 83 observations is achieved for rotation 2136.
We excluded the vector synoptic maps of Carrington rotations 2092, 2099, 2107, 2127, 2139, 2152 - 2155, 2163, 2164, 2166, and 2167 from our analysis since  these maps are less reliable due to low data coverage. 
In most cases these maps are technically full (no empty longitudes), but due to missing observation days, some data points are farther away from the central meridian than normal, which makes the data noisy and partly incorrect.
This is visually seen as varying noise levels and obvious stripes in these maps. 

In addition to the vector field synoptic maps, we also use the more traditional synoptic maps of the pseudo-radial magnetic field $B_r^{ps}$ in order to derive the PFSS magnetic field.
The resolution of high-latitude observations in (sine latitude) synoptic maps defines the highest possible physically meaningful multipole term in PFSS model.
Therefore, we use the higher-resolution synoptic maps of 900 by 1800 pixels for the pseudo-radial field, which allows a sufficiently high multipole expansion without significant artifacts at high latitudes. 
Missing polar regions are not filled in high-resolution pseudo-radial synoptic maps, contrary to the lower resolution (180 by 360) maps.
Pole filling has not been applied to vector field synoptic maps of any resolution.  
This lack of polar values is not critical for the present work where we compare the PFSS model results with those of the vector field. 

To study the orientation of the vector magnetic field we define the inclination, meridional inclination, and azimuth of the magnetic field in the local coordinate system as follows.
Inclination now refers to the magnetic field deviation from the radial line:
\begin{equation}
I = \arctan \left(\frac{\sqrt{B_{\theta}^2+B_{\phi}^2}}{|Br|}\right)
\end{equation}
Inclination varies between 0$^\circ$ and 90$^\circ$ and describes how non-radial the magnetic field is.
It includes no information on the orientation of the field around radial line along either direction.

Meridional (north-south) inclination is the angle between the radial direction (or anti-radial for $B_r < 0$) and the projection of the \textbf{B}-vector to $r - \theta$ -plane
\begin{equation}
I_m = \arctan \left(\frac{B_{\theta}}{Br}\right),
\end{equation}
and varies from -90$^\circ$ to 90$^\circ$. 
It should be noted that  a configuration where $B_r > 0$ and $B_\theta < 0$ gives the same (negative) sign of $I_m$ as a configuration where $B_r < 0$ and $B_\theta > 0$. The
$I_m$ is zero when the field is radial, and a positive (negative) $I_m$ corresponds to a southward (northward) tilt of the field line.
Therefore, a positive (negative) $I_m$ in the northern hemisphere indicates an inclination toward the equator (pole), but a negative (positive) $I_m$ in the south corresponds to an inclination toward the equator (pole).

Following the definitions of the SOLIS/VSM full-disk data, the azimuth (A) is the angle between -$\hat{\theta}$ (northern) direction and the projection of  \textbf{B}-vector to the (horizontal) $\phi - \theta$ -plane. Azimuth is zero in the northward direction and increases counterclockwise

\begin{eqnarray}
A = \arctan \left(\frac{B_{\phi}}{B_{\theta}} \right)  \mbox{,           if } B_{\theta} <  0,\nonumber \\
A = \arctan \left(\frac{B_{\phi}}{B_{\theta}} \right) + 180 ^\circ \mbox{,           if } B_{\theta} >  0.
\label{eq_az}
\end{eqnarray}
According to Equation \ref{eq_az} the azimuth varies from -90$^\circ$ to 270$^\circ$. 
However, we convert the azimuth to vary from 0$^\circ$ to 360$^\circ$ in the further analysis.

It is important to note  that the above definition of inclination is different from that used in SOLIS full-disk data products, where inclination is the angle between \textbf{B} and $\hat{r}$ vectors and varies between 0$^\circ$ and 180$^\circ$.
With the definition used in full-disk data, the longitudinally averaged inclination would be around 90$^\circ$ since inclination would have a clear two-peak distribution with maxima close to 0$^\circ$ and 180$^\circ$.
Since we aim to investigate the long-term evolution of the large-scale field orientation, our definitions must be unambiguous when calculating the longitudinal averages of the inclination and meridional inclination angles.

\subsection{Harmonic expansion of the magnetic field}

If the magnetic field is a potential field in a region where there are no currents present, then it can be expressed in terms of spherical harmonics  \citep{Altschuler_1969,Schatten_1969,Hoeksema_1983, Virtanen_2016}.
The three components of the potential magnetic field are
\begin{equation}
B_r(r,\theta,\phi) = \sum^{\infty}_{n = 1} \sum^{n}_{m = 0}  P^{m}_{n} (\cos \theta)(g^{m}_{n} \cos m \phi+h^{m}_{n} \sin m \phi) C(r,n),
\label{eq:pfss1}
\end{equation}

\begin{equation}
B_\theta(r,\theta,\phi) = -\sum^{\infty}_{n = 1} \sum^{n}_{m = 0}  \frac{\partial P^{m}_{n}}{\partial \theta} (\cos \theta)(g^{m}_{n} \cos m \phi+h^{m}_{n} \sin m \phi) D(r,n),
\label{eq:pfss2}
\end{equation}

\begin{equation}
B_\phi(r,\theta,\phi) = \sum^{\infty}_{n = 1} \sum^{n}_{m = 0} \frac{m P^{m}_{n} (\cos \theta)}{\sin(\theta)}(g^{m}_{n} \sin m \phi - h^{m}_{n} \cos m \phi) D(r,n),
\label{eq:pfss3}
\end{equation}
where $ P^{m}_{n} (\cos \theta)$ are the associated Legendre functions and C(r,n) and D(r,n) are radial functions.
Harmonic coefficients $g^{m}_{n}$ and $h^{m}_{n}$ are derived from the observed photospheric magnetic field, which is the inner boundary condition.

The straightforward solution for radial functions is \citep{Altschuler_1969}
\begin{equation}
C(r,n) = (n+1)\left(\frac{R_s}{r}\right)^{n+2},
\end{equation}
\begin{equation}
D(r,n) = \left(\frac{R_s}{r}\right)^{n+2},
\end{equation}
where $R_s$ is the solar radius. 
However, this solution gives a roughly dipolar magnetic field up to a few solar radii, which is in contradiction with  solar eclipse observations of a roughly radial coronal magnetic field.
Therefore an additional assumption of a radial magnetic field in the coronal source surface ($r_{ss}$, typically 2.5 $R_s$) was implemented.
This leads to the potential field source surface (PFSS) model \citep{Altschuler_1969,Schatten_1969,Hoeksema_1983} with the following radial functions:

\begin{equation}
C(r,n) = \left(\frac{R_s}{r}\right)^{n+2} \left[\frac{n+1+n\left(\frac{r}{r_{ss}}\right)^{2n+1}}{n+1+n\left(\frac{R_s}{r_{ss}}\right)^{2n+1}} \right],
\label{eq:pfss4}
\end{equation}

\begin{equation}
D(r,n) = \left(\frac{R_s}{r}\right)^{n+2} \left[\frac{1 - \left(\frac{r}{r_{ss}}\right)^{2n+1}}{n+1+n\left(\frac{R_s}{r_{ss}}\right)^{2n+1}} \right].
\label{eq:pfss5}
\end{equation}

In this paper we only consider the potential field solution in the photosphere at $r = R_s$, where $C(r,n) \equiv 1$.
The other radial function $D(r,n)$  depends on the $r_{ss}$ at $r = R_s$ only weakly.
We use the value of  $r_{ss} = 2.5R_s$ 

\begin{figure*}[ht]
\begin{center}
\includegraphics[width=\textwidth]{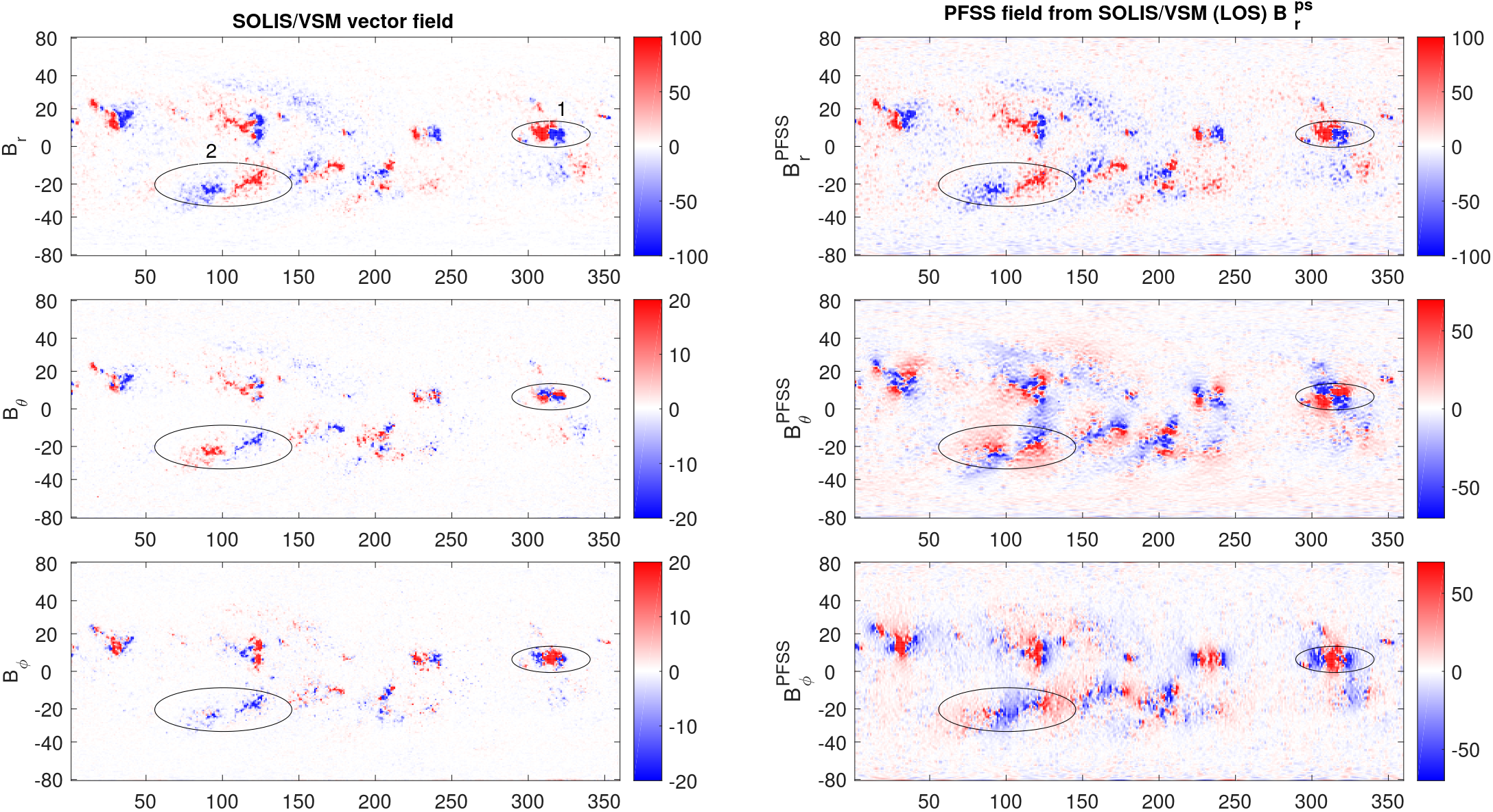}
\caption{Three vector components ($B_r$, $B_\theta$, and $B_\phi$) of the vector field (left) and the potential field (n = 180) approximation (right) from $B_r^{ps}$ for Carrington rotation 2170 in November 2015.The active region in Box 1 includes a sunspot with a strong field, while Box 2 includes a decaying active region. The  color scales in the panels are different.
\label{Fig1}}
\end{center}
\end{figure*}

Harmonic coefficients $ g^{m}_{n}$ and $ h^{m}_{n}$ are solved using the observed photospheric magnetic field (inner boundary condition) as 
\begin{equation}
\left\{ \begin{array}{cc} g^{m}_{n} \\ h^{m}_{n} \end{array} \right\} = \frac{2n+1}{N} \sum^{N_\theta}_{i = 1} \sum^{N_\phi}_{j = 1}  B^{j,i}_{r} P^{m}_{n}(\cos\theta_i) \left\{ \begin{array}{cc} \cos(m \phi_j) \\ \sin(m \phi_j) \end{array} \right\} 
\label{eq:gh}
,\end{equation}
where $B^{j,i}_{r}$ refers to the radial component of the photospheric magnetic field at co-latitude-longitude bin (i,j) and
N refers to the number of data points in the synoptic map. 
In the case of full data coverage each (i,j) bin contains data and $N = N_\phi N_\theta$.
The detailed derivation of Equation \ref{eq:gh} can be found in \cite{wso_doc}.
Equations \ref{eq:pfss1} -- \ref{eq:pfss3} and \ref{eq:pfss4} -- \ref{eq:gh} show how the three vector components of the PFSS field are calculated.
The assumption that the field is potential and that $B_r$ is known in the photosphere also allow also  the harmonic expansion of $B_\theta$ and $B_\phi$ to be solved exactly. 

The accuracy of the PFSS solution increases with increasing n, but the resolution of the input data limits the highest physically meaningful n.
The Nyqvist theorem  applies in longitude direction  requires that $n_{max} < N_\phi/2$, where $N_\phi$ is the number of grid points in longitude, assuming that the spatial resolution of observations is at least the spatial resolution of the synoptic map. 
In the latitudinal direction, the situation is somewhat more complicated, since the data is equally sampled in sine latitude, and therefore the latitudinal width of pixel increases with latitude. 
Legendre polynomial $P^{m}_{n}$ is zero along n$-$m circles of constant latitude. 
In the case of 360 by 180 pixel SOLIS/VSM synoptic maps the width of highest pixel is more than 4$^\circ$, which sets the harmonics beyond n = 20 to be overfitted.
However, the width of pixels decreases rapidly with decreasing latitude.
Moreover, the polar field is roughly unipolar, and increasing the resolution at high latitudes would not necessarily change the harmonics significantly.
Therefore, we use $n_{max} = 180$ in this work, which is sufficient in terms of input data.
However, we note that axial harmonics (m = 0) with large n may suffer from artifacts due to overfitting at the highest latitudes.
The chosen $n_{max}$ is appropriate, also in terms of  presenting the PFSS magnetic field in the same 180 by 360 synoptic grid as the lower resolution synoptic maps of VSM vector field.

Many models of space weather and heliospheric magnetic fields rely on the PFSS model. 
The radial component of the magnetic field in Equation \ref{eq:gh} is often the $B_r^{ps}$ derived from $B_r^{LOS}$ under the assumption of radial magnetic field in the photosphere.
We here compare the PFSS field $B^{PFSS}$ based on $B_r^{ps}$ from $B_r^{LOS}$ with the true vector field $B^{vec}$.

\section{Vector magnetic field}

Figure \ref{Fig1} provides an example of the SOLIS/VSM synoptic maps of the vector field for Carrington rotation 2170 in November 2015. 
Panels on the left from top to bottom show radial $B_r$, meridional $B_\theta$, and zonal $B_\phi$ components of the measured vector field.
Panels on the right show the same components derived from the corresponding PFSS solution based on LOS observations. 
Before delving into the discussion of the topology of large-scale and global magnetic fields, let us validate the orientation of the vector field for a few configurations.
The region inside Box 1 in Figure \ref{Fig1} is an active region with large field intensity that also includes a sunspot.
Box 2 in Figure \ref{Fig1} shows a decaying active region which appears as a plage in chromospheric observations. 
Overall, the orientation of the radial magnetic field component in active regions follows the known tilt and polarity patterns characteristic to solar cycle 24.
According to the Hale polarity rule \citep{Hale_1919,Bruzek.Durrant1977} the leading (trailing) polarity part of bipolar active regions (in cycle 24) is typically negative (positive) in the north and positive (negative) in the south.

The observed radial and meridional fields (see Fig. \ref{Fig1}) show a systematic pattern where $B_r$ and $B_\theta$ of leading flux of decaying active regions have the same sign in the northern hemisphere, but are oppositely signed in the south (e.g.,  decaying AR at the south at about 235$^\circ$ longitude, 5$^\circ$ latitude).
Zonal field $B_\phi$ is systematically negative in Box 2.

\begin{figure}[tpbh]
\begin{center}
\includegraphics[width=\columnwidth]{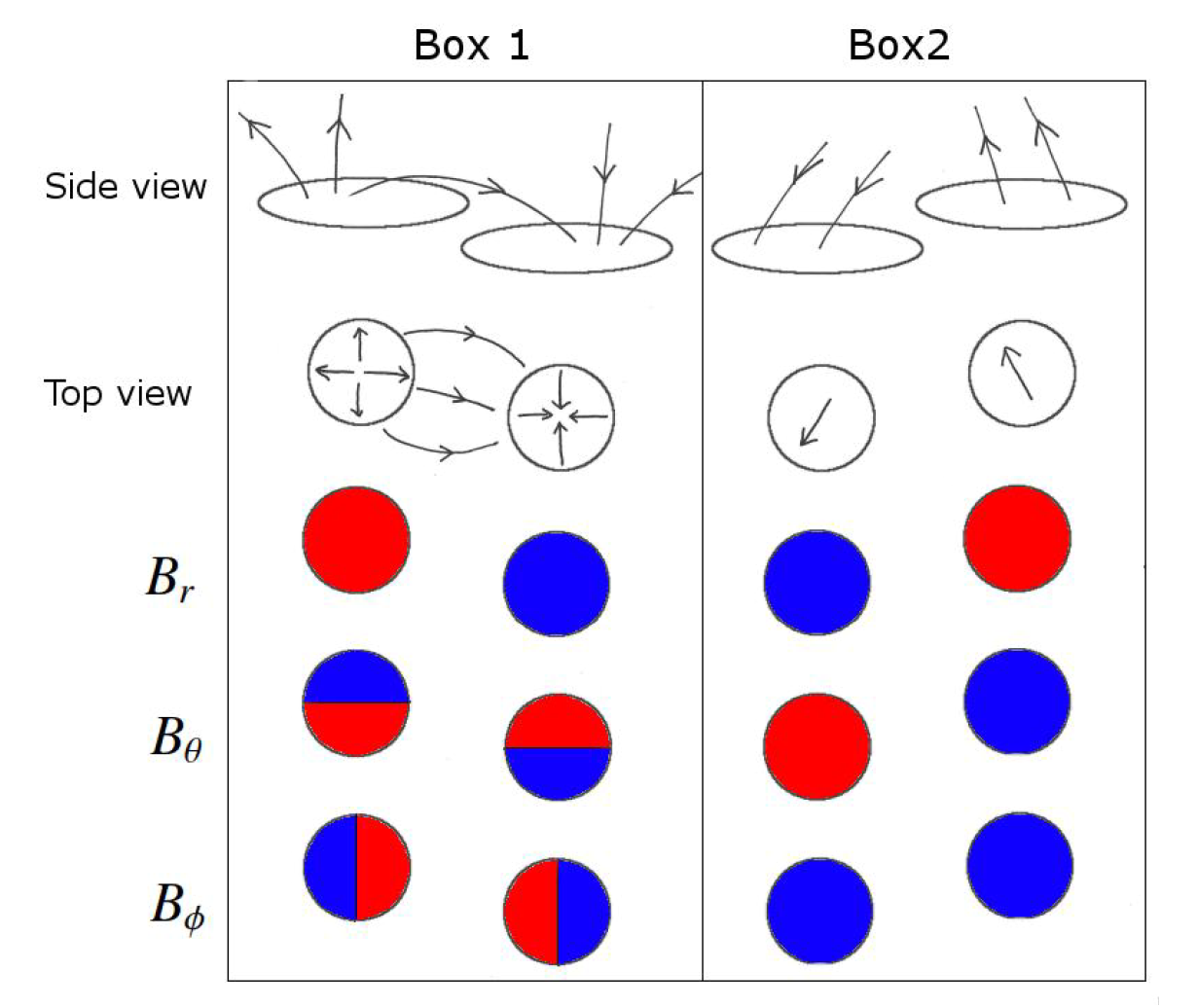}
\caption{Sketch of magnetic field line configuration and related polarity patterns for the active region (Box 1) and for the decaying active region (Box 2) in Figure \ref{Fig1}. Shown are the positive component (red) and  the  negative component (blue). 
}
\label{Fig2new}
\end{center}
\end{figure}

Figure \ref{Fig2new} provides a graphical explanation for the observed sign-patterns for the three components of the vector field for Boxes 1 and 2 in Figure \ref{Fig1}. 
The upper panels of Figure \ref{Fig2new} illustrate the side view, and the second panels the top view of active regions inside Box 1 and Box 2.
The three color-coded bottom panels show the polarity structure of $B_r$, $B_\theta$, and $B_ \phi$  for field line configurations depicted above.
The strong active region in Box 1 has a configuration where the magnetic field expands super-radially from both polarities of the bipolar active region. 
Two circles with radially directed vectors (top view of Fig. \ref{Fig2new}) represent sunspots of opposite polarity, and the arches between them depict closed field lines connecting these two sunspots. 
In terms of polarities of vector components, this is related to the configuration where $B_r$ is   positive in the left pole of the bipole and negative in the right pole. 
Due to the super-radial expansion the $B_\theta$ component is negative (positive) in the northern (southern) part of the left side pole, and oppositely oriented in the right side pole. 
Similarly, $B_ \phi$ is negative (positive) at the left (right) end of the left side pole and oppositely oriented in the right side pole.
The observed polarity structures in Box 1 of Figure \ref{Fig1} correspond very well to the idealized polarity configuration in the left panels of Figure \ref{Fig2new}.

The $B_r$ component in Box 2 of Figure \ref{Fig2new} is negative in the left pole of the bipole and positive in the right pole.
This corresponds to the Hale law.
Moreover, the decaying active region also depicts  Joy's law with the leading flux laying at a clearly lower latitude. 
The $B_\theta$ component is positive in the left pole and negative in the right pole of Box 2.
This corresponds to the decaying active region inside Box 2 in Figure \ref{Fig1}.
Box 2 in Figure \ref{Fig2new} depicts a configuration where magnetic field lines in both footpoints are inclined from radial toward the equator, and the systematic negative zonal component shows that the field is closing. 
The overall structure is still arc-shaped, but the plane of the arcade is systematically tilted toward the equator. 
The significant difference between Box 1 and Box 2 of Figure \ref{Fig2new} is that the field does not expand super-radially in Box 2. 

\begin{figure}[tttt]
\begin{center}
\includegraphics[width=\columnwidth]{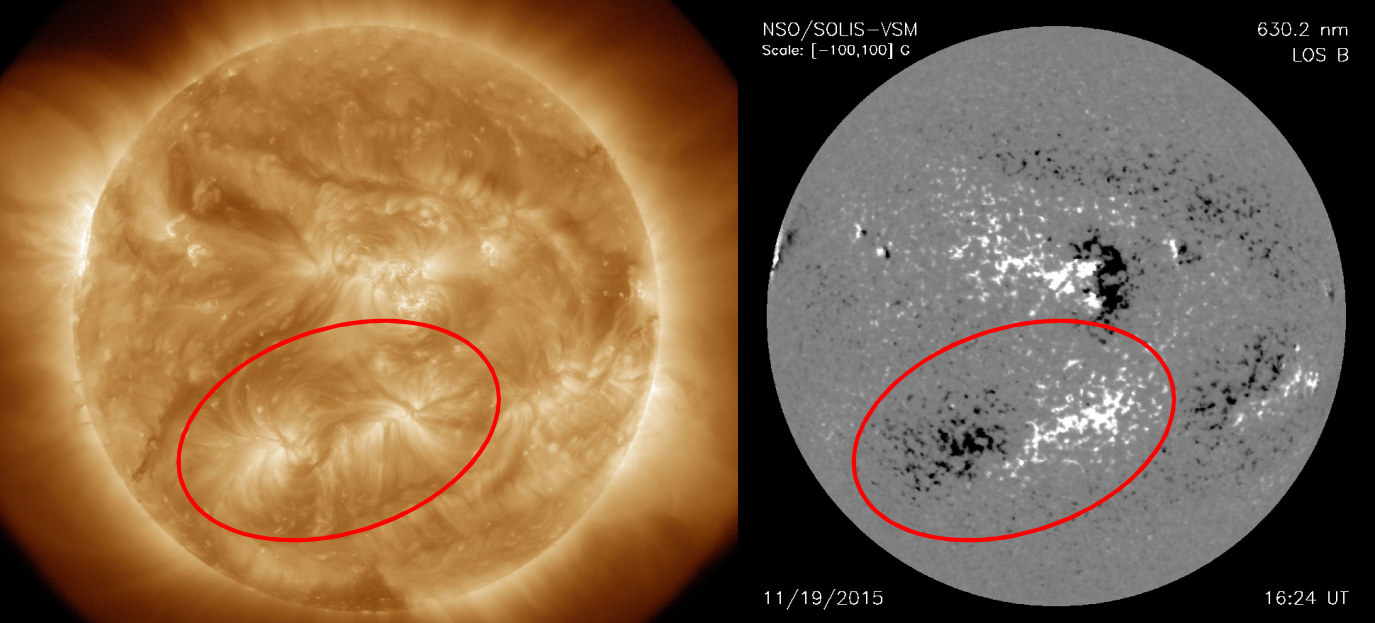}
\caption{AIA 171Å image and corresponding SOLIS/VSM $B_{LOS}$ full-disk magnetogram on November 19, 2015. 
}
\label{Fig_AIA}
\end{center}
\end{figure}

As noted above, the plane of the arcade included in Box 2 is inclined  northward (toward the equator) relative to the local radial direction.
However, photospheric observation alone cannot exactly show how the field is closing between the footpoints and what  the orientation of the arcade is at higher altitudes.
Therefore, we also examine the simultaneous AIA 171Å image, which shows the coronal magnetic loop configuration.
Figure \ref{Fig_AIA} shows the AIA 171Å image and the corresponding SOLIS/VSM $B_{LOS}$ full-disk magnetogram on November 19, 2015, when the decaying active region of Box 2 (marked by the red oval in Figure \ref{Fig_AIA}) was close to central meridian. 
Both footpoints are clearly seen as bright regions in the AIA image and there are visible arcades connecting the polarities. 
We note that the image quality might not be  sufficient to show all the details here, and that it  appears  in more detail  in the solar monitor.\footnote{\url{https://www.solarmonitor.org/data/2015/11/19/pngs/saia/saia_00193_fd_20151119_214853.png}} 
The AIA image shows that the plane of the arcade is inclined northward especially close to the leading polarity region.
However, the inclination angle of the plane from the local radial direction cannot be exactly defined based on observations from a single vantage point, and therefore the AIA image can only partially verify the pattern of Box 2.
If the field lines of an arcade at mid-latitudes of the southern hemisphere are seen to be oriented northward in this LOS projection, we can be sure that the plane of this arcade is inclined northward from radial. 
On the contrary, if the field lines of another arcade are oriented southward in the LOS projection, we cannot determine whether the arcade is inclined north or south from radial.
There is also some indication of twisting of the plane toward the trailing polarity region.  It should be noted  that this discussion refers to one particular loop that we selected to demonstrate the validity of the derivation of the topology based on vector data. 
Not all loops on the Sun should be expected to exhibit a similar topology.

\subsection{Comparison between vector and PFSS fields}

We now compare the components of the vector magnetic field (Fig. \ref{Fig1}, left panels) with those derived from PFSS extrapolation based on B$_r^{ps}$ input (Fig. \ref{Fig1}, right panels). 
The polarity structure of all  three components of the active region in Box 1 is similar for the vector field and the PFSS field and corresponds to the sketch in the left panels of Figure \ref{Fig2new}.
We note that the active region in Box 1 is considerably larger in $B_\theta^{PFSS}$ and $B_\phi^{PFSS}$ than in vector $B_\theta$ and $B_\phi$, but there is no considerable difference between B$_r^{PFSS}$ and B$_r$.
Thus, the observed vector magnetic field and the PFSS field show a very similar structure for the active region of Box 1 despite the difference in areas and some other minor differences.
This implies that the active region fields are mainly potential and the effect of plasma flows and currents is minor.

The decaying active region of Box 2 also shows a similar polarity configuration for vector $B_r$ and $B_r^{PFSS}$.
However, the vector field  $B_\theta$ depicts systematic positive polarity in the left footpoint from $-$15$^\circ$ to $-$30$^\circ$  latitude, but $B_\theta^{PFSS}$ depicts positive polarity from about $-$10$^\circ$ to $-$25$^\circ$, and negative polarity from $-$25$^\circ$ to $-$40$^\circ$ latitude.
The right footpoint shows a negative unipolar structure in $B_\theta$ from about $-$10$^\circ$ to $-$20$^\circ$, but $B_\theta^{PFSS}$ reverses sign at about  $-$15$^\circ$.
A notable difference is also seen between the vector $B_\phi$ and $B_\phi^{PFSS}$.
Vector $B_\phi$ is continuously negative, while  $B_\phi^{PFSS}$ reverses sign at the centers of the two footpoints, being negative between the footpoint centers and positive outside them.

Accordingly, the vector field $B_\theta$ and $B_\phi$ show a simpler, more unipolar structure  than $B_\theta^{PFSS}$ and $B_\phi^{PFSS}$ for the decaying active region in Box 2.
This can be understood by considering the following picture. 
Let us imagine a simple magnetic arcade connecting the two ends of a fluxrope. 
In the absence of gas pressure, the field originating from each footpoint will expand significantly with height, developing a pattern, in which the magnetic field vectors anchored in the opposite polarity ends will be inclined in the direction away from each other.
The left panels of Fig. \ref{Fig2new} depict this type of structure for an active region.
This overexpansion is clearly seen to be the pattern of $B_\theta^{PFSS}$ and $B_\phi^{PFSS}$  not only for the active region in Box 1, but also for the decaying region in Box 2.
Contrary to the latter, the pattern of vector field $B_\theta$ and $B_\phi$ shows that magnetic vectors in the opposite footpoints of the arcade are inclined toward each other, which corresponds to arcade without over-expansion.

Most of the photosphere is covered by weak fields of the background magnetic network.
This network is mainly governed by supergranulation, which has a typical lifetime of about 20 hours. 
Due to this short lifetime, synoptic maps do not represent the magnetic network in detail and we cannot expect the magnetic field models to reproduce the network structure very well.
The magnitude of active region $B_r^{PFSS}$ is about the same as vector field $B_r$, except that the background magnetic network  $B_r^{PFSS}$ shows slightly higher values than $B_r$ 
(this can be also seen in Figure \ref{Fig1} in the more colorful plot of $B_r^{PFSS}$).
However, the magnitudes of $B_\theta^{PFSS}$ and $B_\phi^{PFSS}$ are much larger than those of the  $B_\theta$ and $B_\phi$. 
We note the different color scales in Fig. \ref{Fig1}.
This suggests that the vector magnetic field is considerably more radial than the PFSS field. 
This difference is partly due to the different field strengths in $B_r^{ps}$ and vector $B_r$ because of the effect of different filling factors. 
For vector data the filling factor is derived from spectral line profiles, being close to unity in sunspots, about 0.15 in plages, and even smaller in areas outside plages. 
The LOS $B_r^{ps}$ field is derived from spectral line  profiles using the center of gravity method \citep{Rees_1979}, where the filling factor is set to unity everywhere.
The filling factor describes the fraction of magnetized  plasma in the observed region (pixel) (assuming that a part of the plasma is not magnetized at all and the rest of the area is uniformly magnetized).

The observed value of field intensity is multiplied by the filling factor when deriving the correct average magnetic flux density of the pixel.
Therefore, applying the filling factor makes the background magnetic field intensities of the vector field smaller than the  $B_r^{ps}$ and PFSS fields.

These results on the vector field versus PFSS comparison indicate that $B_\theta^{PFSS}$ and $B_\phi^{PFSS}$ are reliable only in active regions with large field intensities.
In the decaying active regions the magnetic field deviates more from potential field configuration due to the increasing effect of plasma.
Decaying active regions have excessively large $B_\theta^{PFSS}$ and $B_\phi^{PFSS}$ because the lack of plasma in the PFSS model makes the field overexpand and become less radial.

\section{Large-scale structure of the vector magnetic field in 2010 - 2017}

\begin{figure*}[t]
\begin{center}
\includegraphics[width=\textwidth]{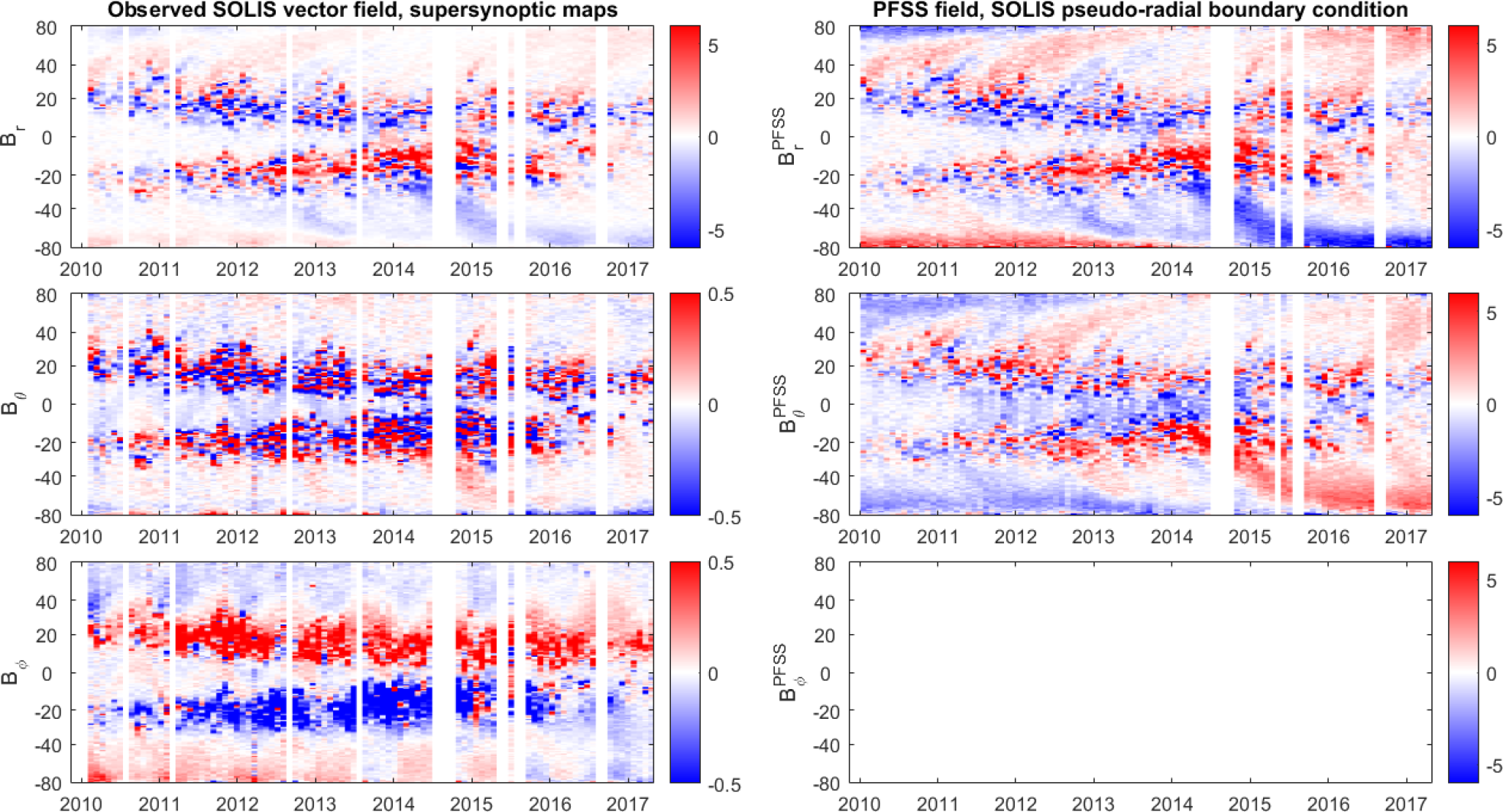}
\caption{Supersynoptic map of radial $B_r$, meridional  $B_\theta$, and zonal $B_\phi$ SOLIS/VSM vector observations (left) and PFSS solution based on SOLIS/VSM 
$B_{LOS}$ observations (right). The  color scales in the panels are different.
\label{Fig2}}
\end{center}
\end{figure*}

Figure \ref{Fig2} shows the longitudinal averages of the synoptic maps in 2010 - 2017, depicting the three components of the photospheric magnetic field vector observed by SOLIS/VSM and derived using the potential field model and LOS $B_r^{ps}$.
This presentation is often called the ``magnetic butterfly'' diagram, following the terminology first used for sunspots, or the supersynoptic map of the magnetic field. 
We use the latter term in this paper.
It should be noted  that \cite{Ulrich_2002} used the term supersynoptic when referring to a map where synoptic maps are only squeezed in longitude without averaging, while supersynoptic maps in this paper show rotational longitudinal averages.

The vector radial magnetic field $B_r$  and the PFSS field $B_r^{PFSS}$ (upper panels of Figure \ref{Fig2}) show the well-known solar cycle evolution of the solar magnetic field.
The mid-latitude belts in $B_r$ show the mixed polarity pattern of the bipolar active regions.
Poleward surges of opposite polarity in the two hemispheres and the approximate timings for polar field reversal 
can be identified in Figure \ref{Fig2}.
The radial fields of active regions are quite similar for vector and PFSS fields, but the magnitude of the high-latitude field is considerably larger for $B_r^{PFSS}$.
This can be explained by the  filling  factors.

The topology of $B_r$ and $B_r^{PFSS}$ fields agree very well, in agreement with the sample depicted in Figure \ref{Fig1}.
The annual variation due to the vantage point effect \citep[see, e.g.,][and references therein]{Virtanen_2017A} appears in the vector $B_r$ already around 60$^\circ$ latitude, but it is not visible in the PFSS solution.
It should be noted that the relative contribution of transverse fields (i.e., transverse to the line of sight in plane of sky coordinates)  to $B_r$ increases toward high latitudes. 
Since this component is noisier than the LOS component, annual variation in the viewing angle of high-latitude areas results in the modulation of the noise level, and reveals itself to be a stronger annual variation in $B_r$ than in $B_r^{PFSS}$.

With the exception of one surge, the vector field B$_\theta$ and B$_\phi$ components are too noisy to distinguish any patterns associated with surges. 
The surge of negative B$_r$ that starts in the southern hemisphere before the end of 2014 and continues until early 2016 shows a systematic pattern of positive B$_\theta$.
This pattern suggests that the magnetic field in this surge is inclined toward the equator.
Outside active regions $B_\theta$ is rather noisy, but $B_\theta^{PFSS}$  indicates a clearly structured polarity pattern.
At about 30$^\circ$ - 60$^\circ$, B$_\theta$ shows roughly the same overall polarity pattern as $B_\theta^{PFSS}$ in both hemispheres.
The general pattern for most of the time depicted in Fig. \ref{Fig2} is that $B_r$/$B_r^{PFSS}$ and $B_\theta$/$B_\theta^{PFSS}$ have same signs in the north and opposite signs in the south.  
This pattern implies that, on average, the magnetic field vector in areas outside of active regions is inclined toward the solar equator.
\cite{Ulrich_2013} found the field to be inclined toward the poles, but their study considered only highest latitudes above $\pm 85^\circ$.  

The meridional component shows an interesting temporal evolution between 2010 and 2014. 
There is a narrow band of negative $B_\theta$ between 60$^\circ$ and 75$^\circ$ of southern latitude, which appears stronger and broader in  $B_\theta^{PFSS}$. 
This band slowly drifts toward the south, and by early 2014 reaches the polar latitudes.
This is roughly the time of south pole reversal \citep[28 October 2013; see Table 1 in][]{Mordvinov.etal2016}. 
The location of this band also corresponds to the location of the polar crown filaments prior to polar field reversal. The sign of 
$B_\theta$ is systematically negative in this band, while $B_r$ is positive, reflecting again the equatorward inclination of the magnetic field.  
A similar development of negative $B_\theta$ is seen in northern hemisphere high latitudes in 2010 - 2011 in vector $B_\theta$ and in 2010 - 2012 in $B_\theta^{PFSS}$, although the band is more narrow and the reversal is earlier than in the south.

The zonal component  of the active region in Fig. \ref{Fig2} is positive (westward) in the northern hemisphere and negative (eastward) in the southern hemisphere, reflecting the Hale polarity rule for solar cycle 24 (see Fig. \ref{Fig2new}).
Longitudinal averages of the vector $B_\phi$ show a significant annual variation and almost zero values at the poleward side of active regions, and at the equator.
Outside the active regions, $B_\phi$ is typically negative in the north and positive in the south until mid-2015.
Thereafter $B_\phi$ in the north is very weak and $B_\phi$ in the south is systematically negative.
\cite{Ulrich_2005} and \cite{Lo_2010} derived the zonal field from LOS observations and found that $B_\phi$ poleward of active regions has polarity opposite to that of  $B_\phi$ in the  active region after the maximum of the solar cycle, corresponding to the start of the new zonal cycle.
Our results show that $B_\phi$ poleward of mid-latitudes has a sign opposite to that of the active region $B_\phi$ from the ascending phase to the  early declining phase of the solar cycle.
\cite{Pipin_2014} derived $B_\phi$  from MDI LOS observations and found oppositely signed $B_\phi$ in the active regions and high latitudes from 1996 - 2009  in the early ascending phase of solar cycle 23.
Their results show that the high-latitude zonal field was already negative in the south  in 1997, close to solar cycle 22 minimum, which contradicts \cite{Ulrich_2005} and \cite{Lo_2010}.
However, \cite{Pipin_2014} found that $B_\phi$ was systematically positive (negative) in the north (south) from 2003 to 2010, which agrees with  \cite{Ulrich_2005} and \cite{Lo_2010}, but disagrees with our results for the year 2010.
The results presented here on the zonal component based on the vector field measurements obviously do not follow the LOS-based results of the zonal cycle.
Further studies are needed to clarify these differences and to understand the limitations in vector $B_\phi$ and $B_\phi$ derived from LOS data.

It should be noted that the PFSS solution (Equations \ref{eq:pfss1} - \ref{eq:pfss3}) depends on $\phi$ only via the $\sin m \phi$ and $\cos m \phi$ terms.
Since m is an integer, the longitudinal averages of the PFSS field vanish except for the axial (m = 0) terms.
Moreover, since $B_\phi$ (Eg. \ref{eq:pfss3}) is multiplied by m, even the axial term contribution to $B_\phi$ vanishes. 
Therefore, the longitudinal average of $B_\phi^{PFSS}$ is exactly zero for all times (see Fig. \ref{Fig2}).

\subsection{Polarity match between vector and PFSS field}

\begin{figure}[tpbh]
\begin{center}
\includegraphics[width=\columnwidth]{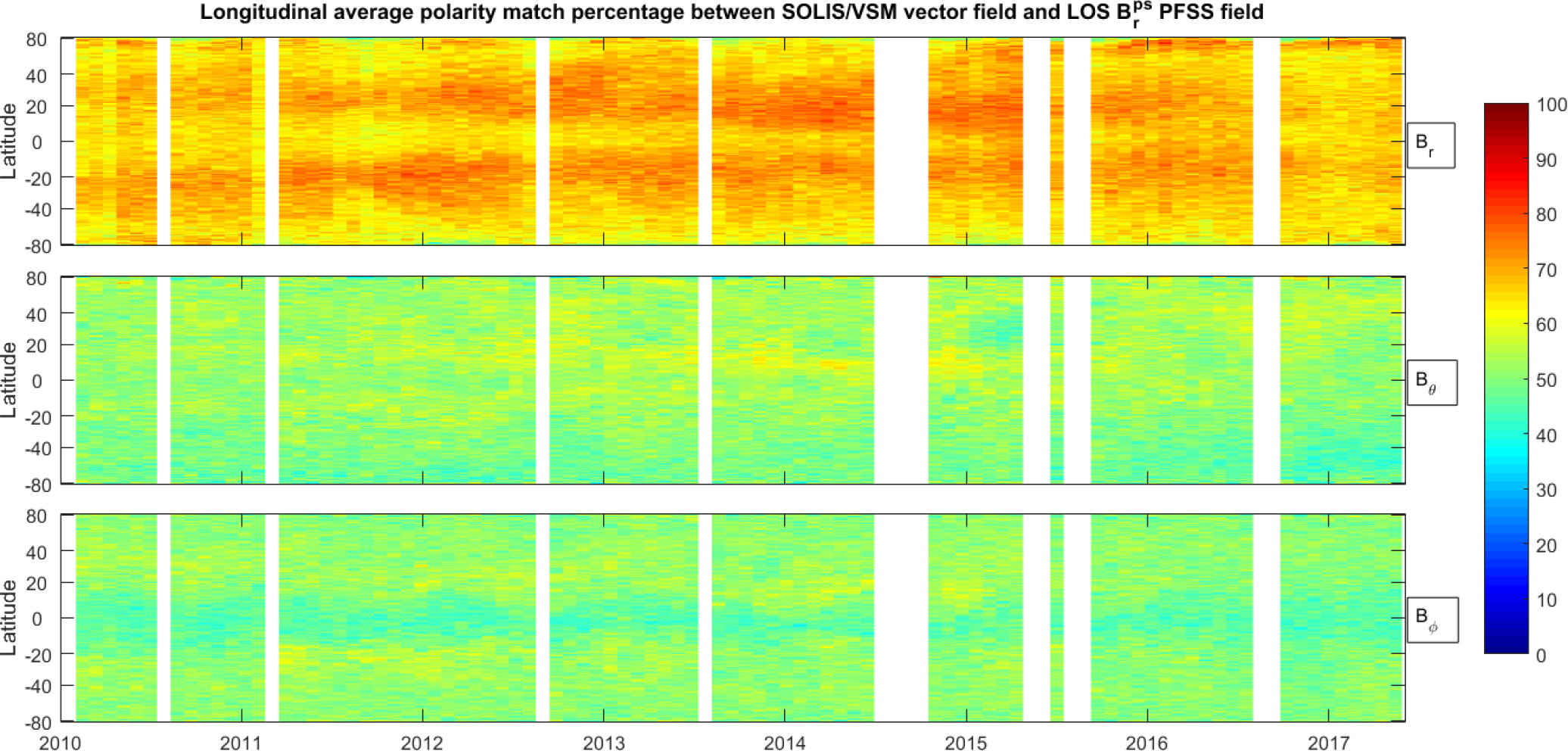}
\caption{Pixel-by-pixel polarity match percentages between the vector magnetic 
field and the PFSS field. 
The three panels show the  match percentages for $B_r$, $B_\theta$, and $B_\phi$.
\label{Fig2b}}
\end{center}
\end{figure}

Figures  \ref{Fig1} and \ref{Fig2} indicate that the polarity structure of the radial component  $B_r^{PFSS}$ of the potential field corresponds to the vector field $B_r$ very well, but the 
agreement between vector meridional and zonal components with the corresponding PFSS components is weaker.
Figure \ref{Fig2b} shows that the rotational polarity match percentages for the three components of the vector field and PFSS solution.
Signs of the radial, meridional, and zonal components of the vector and PFSS fields are compared to each other pixel-by-pixel for each 360 by180 synoptic map and the percentage of pixels with same signs in observed and PFSS field are calculated along the longitude for each latitude bin.
The vertical columns in Figure \ref{Fig2b} are color-coded and show the rotational polarity match percentages.
Figure \ref{Fig2b} shows that more than 80\,\% of the pixels of the radial PFSS field give the same polarity around active regions as the vector radial field.
Outside active regions the match is weaker, but still typically more than 60\,\%.
At high latitudes the vector field maps are very noisy, which decreases the match percentages.
The overall polarity match for the radial component is about 70\,\%.

The agreement between the meridional components $B_\theta$ and $B_\theta^{PFSS}$ is quite random, typically 50\,\%, and  at best only about 60\%.
The polarity match between the zonal components ($B_\phi$  and $B_\phi^{PFSS}$) is typically below 50\% at the equator, which suggests  systematically opposite polarities in this region. 
However, it should be noted that the zonal field is weak and non-systematic at the equator. 
Disagreement of $B_\phi$  and $B_\phi^{PFSS}$ may relate to the dominant effect of noise in the observations.
The best match of about 60\% for the zonal field is found in active regions, but  higher latitude fields also show match values that are slightly above  50\%.

\section{Inclination, meridional inclination, and azimuth of the vector magnetic field}

\begin{figure*}[hhhhht]
\begin{center}
\includegraphics[width=\textwidth]{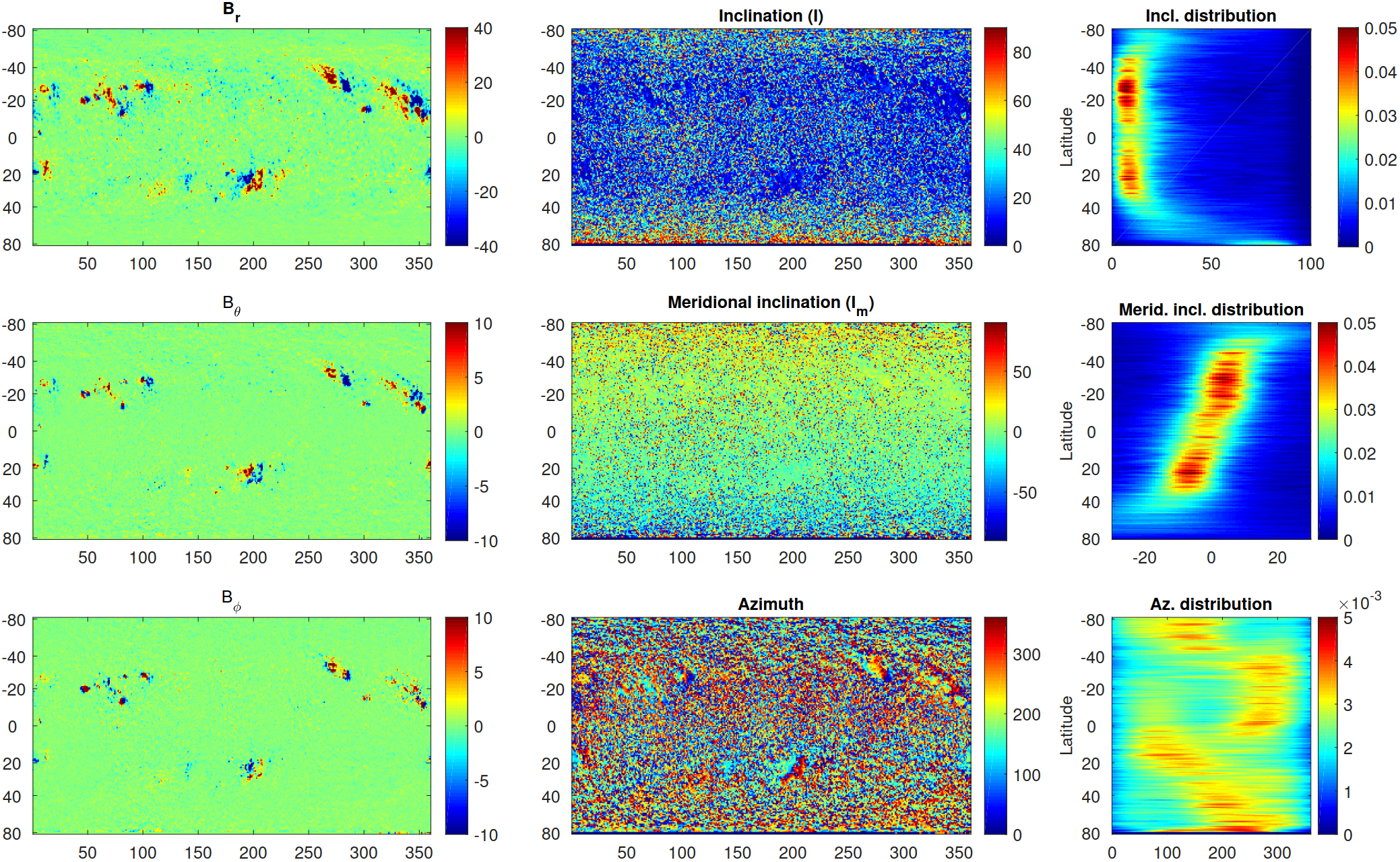}
\caption{Left: Synoptic maps of the three components of the vector magnetic field observed during CR 2100 (August 2010).
Middle: Corresponding synoptic maps of inclination from radial, meridional inclination, and azimuth.
Right: Distribution of the values of inclination from radial, meridional inclination, and azimuth for each latitude bin during CR 2100. The color-coding in every horizontal row gives the histogram of values in the corresponding latitude bin. All histograms are normalized to one.
\label{Fig3}}
\end{center}
\end{figure*}

The left  panels in  Figure \ref{Fig3} show the synoptic maps of the three components of the  vector magnetic field for CR 2100, August 2010.
We note that the color map is now different from  Figure \ref{Fig1} in order to make small-scale features more visible.
The middle panels show the inclination, meridional inclination, and azimuth for every pixel in the synoptic map.
The right  panels are color-coded to  show the distribution of values of inclination, meridional inclination, and azimuth for each latitude bin. 

Inclination of the field in the active regions is typically rather small, often close to zero, indicating that the strong fields are close to radial in the photosphere  (at  1 degree by 1 degree pixel size of synoptic maps).
The inclination increases toward the poles, and its distribution systematically shifts to higher values with increasing latitude. 
However, the results poleward of $\pm 60^\circ$ are probably  questionable due to increasing noise, as discussed above.

By definition, meridional inclination is always smaller than total inclination.
The histogram distribution depicted in Figure \ref{Fig3} shows clearly that the meridional inclination increases with latitude fairly systematically. 
Meridional inclination is mainly negative in the south and positive in the north, showing that the field is typically inclined toward the equator.

The average distribution of the azimuth is less regular, but it appears to favor four azimuthal directions: 90$^\circ$, 150$^\circ$, 210$^\circ$, and 270$^\circ$.
The azimuth in the northern active regions up to about $30^\circ$ is typically about  $270^\circ$ (westward) and the corresponding southern latitudes about $90^\circ$ (eastward).
This reflects the structure shown in Figure \ref{Fig2}, where active region $B_\phi$ is systematically positive in the north and negative in the south. 
At northern high latitudes of about $40^\circ$ - $60^\circ$ the azimuth is centered at about $150^\circ$. 
In the south the azimuth changes more smoothly with latitude and  the value of  $210^\circ$ is reached at about $-50^\circ$ latitude.

\subsection{Evolution of inclination and meridional inclination}
\begin{figure*}
\begin{center}
\includegraphics[width=\textwidth]{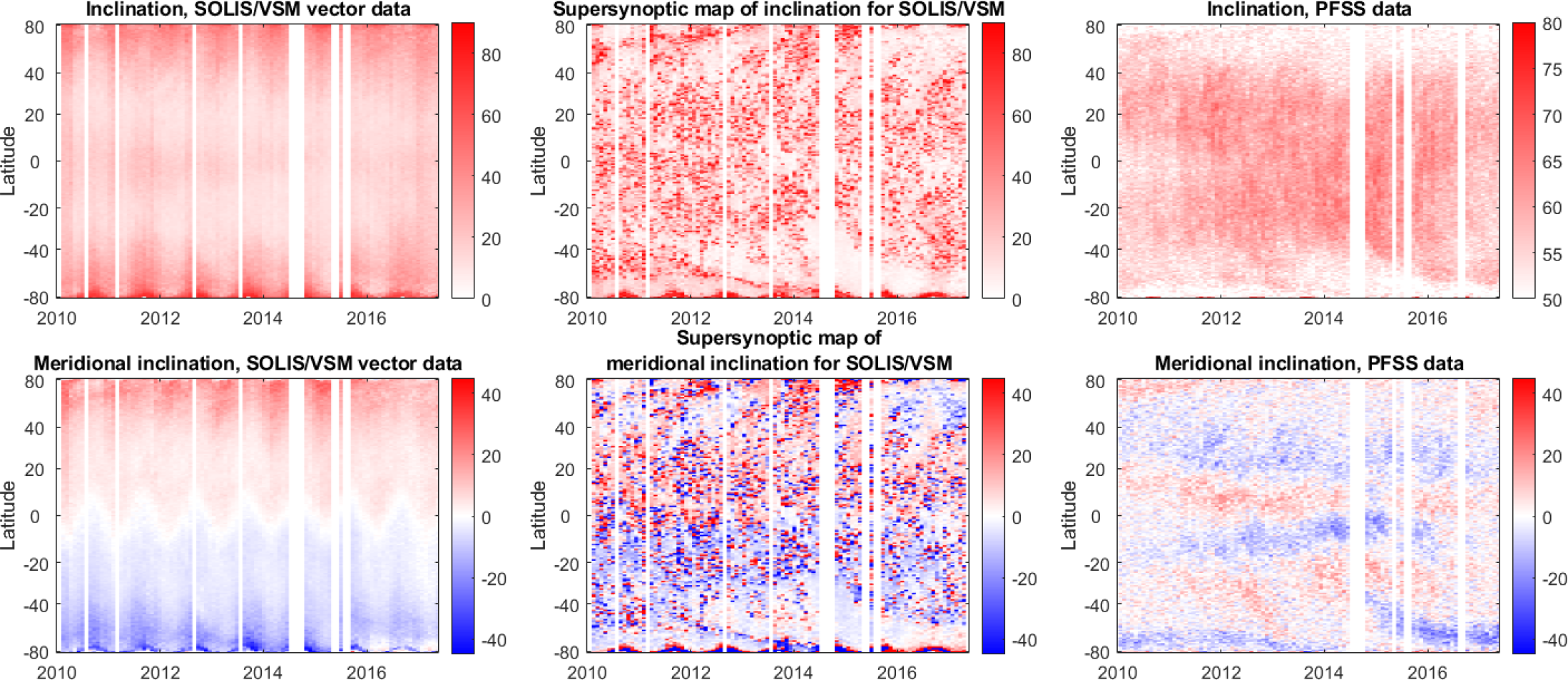}
\caption{Left:  Supersynoptic mean inclination from radial and meridional inclination for SOLIS/VSM vector data. 
Middle: Inclination from radial and meridional inclination derived from supersynoptic (longitudinally averaged, shown in Fig. \ref{Fig2}) SOLIS/VSM data.
Right: Supersynoptic mean inclination and meridional inclination for PFSS data. 
\label{Fig4}}
\end{center}
\end{figure*}

Figure \ref{Fig3} shows that the distribution of inclination and meridional inclination of the photospheric magnetic field has a well-defined maximum for each latitude bin, and that these maxima, in most cases, vary fairly systematically with latitude. 
A visual inspection of all rotations verifies that this pattern is persistent. 
Also, the distribution of azimuth has a maximum for each latitude bin, but the location of the maximum varies from one cycle to another and the latitudinal patterns are less systematic.
Therefore, we leave the topic of the long-term evolution of azimuth to subsequent studies. 
Figure \ref{Fig4}, left panels, presents the supersynoptic maps of inclination and meridional inclination for vector magnetic field.
Each column shows the rotational mean value over longitude for all latitudes. 
The middle panels show the inclination and meridional inclination derived from sypersynoptic maps (longitudinal averages) of SOLIS/VSM vector data (shown in Fig. \ref{Fig2}).
The panels on the right show supersynoptic maps of inclination and meridional inclination for PFSS model field.

Figure \ref{Fig4}  proves that the mean inclination does not have a significant solar cycle evolution; instead,  the pattern of mean inclination is quite systematic and persistent.
The inclination is smallest at low latitudes, in the region of strongest magnetic fields, and increases toward the poles. 
There is also a smaller peak at the equator. 
Annual variation is surprisingly large in vector field mean inclination and is seen not only at high latitudes, but already at about $\pm 30^\circ$   latitude.
The amplitude of the annual variation is larger in the south and has there its maximum and minimum in the September and March, respectively.

The mean meridional inclination of SOLIS/VSM (lower left panel in Fig. \ref{Fig4} ) also shows quite a persistent structure. 
Meridional inclination is around zero at the equator and increases with latitude.
The sign of meridional inclination reflects the configuration where magnetic field lines are inclined toward the equator.
Annual variation in mean meridional inclination is seen already at the equator. 
It seems that the field appears most radial close to the latitude of the Earth's vantage point, since the zero value of mean meridional inclination varies roughly in phase with the Earth's heliographic latitude.
Potentially, this will introduce an annual modulation in signal-to-noise ratio, and could cause the annual pattern in mean meridional inclination shown in Figure \ref{Fig4}.

Inclination of the supersynoptic (longitudinally averaged) field differs significantly from the all field mean inclination.
Active regions appear  more inclined there.
Also,  mean inclination does not systematically vary with latitude, but there are regions of very small inclination over the entire latitude range.
Strong surges of magnetic flux, especially the one in the south in 2014-2016, have a very small mean inclination.

Meridional inclination patterns are also less systematic when the angle is derived from longitudinally averaged rotational values of the magnetic field (supersynoptic map).
Dominant red in the north and blue in the south indicate that field is mainly inclined toward the equator, but there are significant variations especially in active regions. 
Mean meridional inclination appears weaker in unipolar surges of magnetic flux, especially during the strong surge of the south. 
There is also some weakly poleward tilted field in the later phase of this surge.
Interestingly, the poleward edges of surges, especially  the one that starts at about $50^\circ$ north in late 2012 and reaches the pole in 2014 (see Fig. \ref{Fig4}), appear more tilted than the high-latitude field in general.
This boundary also corresponds to a filament channel of polar crown filaments, which are related to polar field reversal \citep{Gopalswamy_2016}.
Annual variation is limited to a smaller latitude range (only around the poles) when deriving inclination and meridional inclination from rotational means. 
This indicates that the large annual variation seen in mean values is a consequence of noisy weak fields.

The PFSS model gives quite different results for inclination (upper right panel of Fig. \ref{Fig4}) than the vector field observations.
The PFSS inclination does not have any significant latitudinal or temporal variation. 
The typical value of PFSS inclination is surprisingly large, about $60^\circ$, and low-latitude fields are slightly more inclined than polar fields.
This shows, according to the PFSS model, that there is a large amount of magnetic field that is already closing  at low altitudes.

Meridional inclination of the PFSS field (lower right panel of Fig. \ref{Fig4})  is also different from the corresponding vector field.
The highest latitudes follow the same pattern of equatorward inclination as the vector field, but the angle is smaller. 
However, both hemispheres have a region of poleward inclined field between latitudes $\pm 20^\circ$ and $\pm 60^\circ$.
This region widens in time toward the equator in both hemispheres, and also toward the pole in the south. 
In the south we also see a region of northward inclined field proceeding poleward from 2014 onward.
This is related to the surge of new flux with negative polarity, which significantly intensified the southern polar field of SC 24 (see also Fig. \ref{Fig2}).
The different latitudinal profiles of meridional inclination between vector and PFSS data most likely relate to the current-free approximation of PFSS model.

\section{Average inclination and meridional inclination}

Since Figure \ref{Fig4}  showed that neither the inclination nor the meridional inclination have a significant solar cycle evolution, it is justified to calculate their latitudinal averages over the entire data set, covering almost one solar cycle.

\begin{figure}[tpbh]
\begin{center}
\includegraphics[width=\columnwidth]{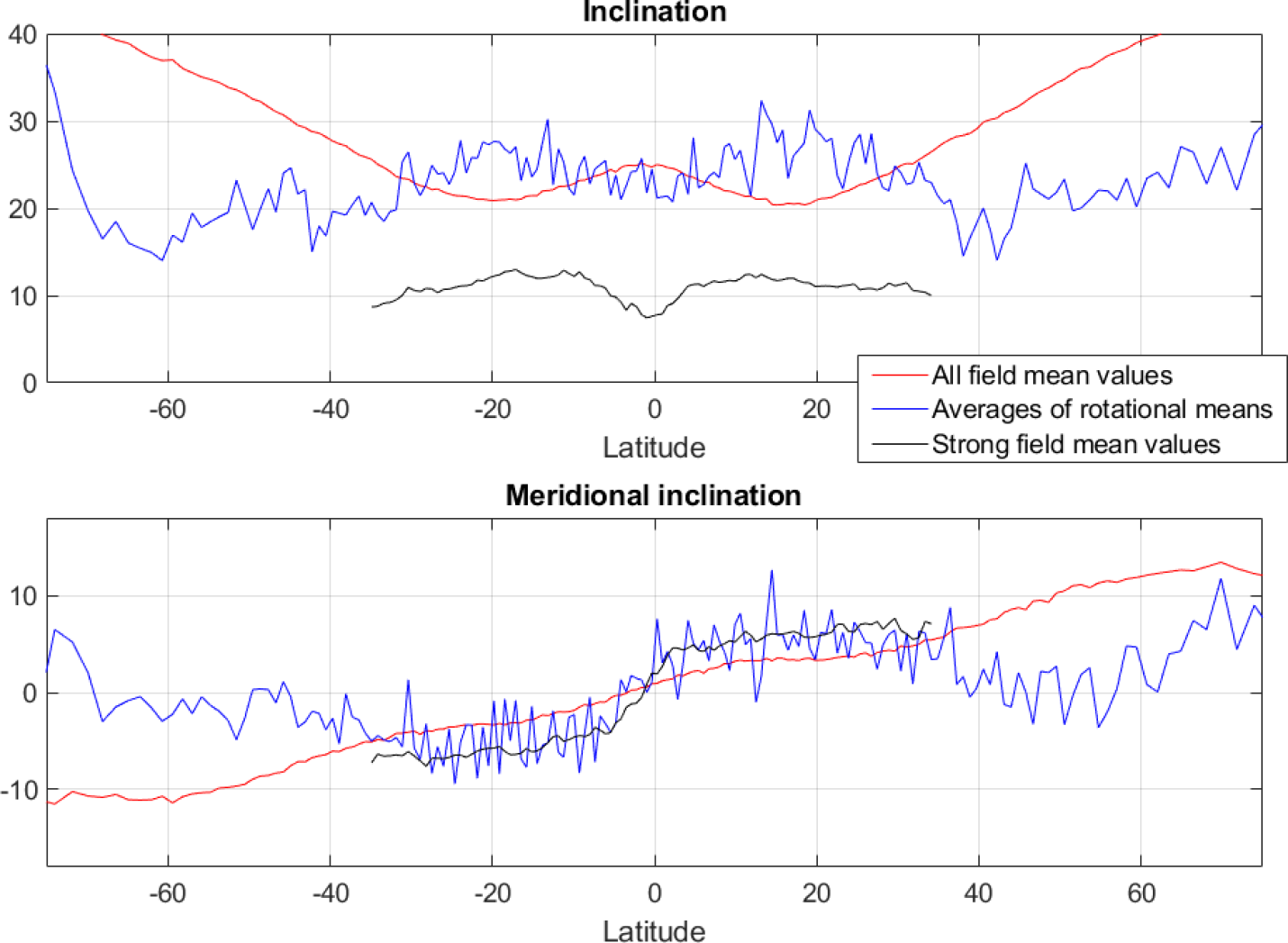}
\caption{Top: Inclination from radial; bottom: meridional inclination of the magnetic field. Red line: mean of all values, black line: mean of strong fields, blue line: means of rotational means.
\label{Fig6}}
\end{center}
\end{figure}

Figure \ref{Fig6} shows the average inclination and average meridional inclination for each latitude bin between $-70^\circ$ and $70^\circ$ derived using three different methods. 
Pixels with weak magnetic fields may introduce large uncertainties and distort the averages, especially for rotations with a large number of weak field pixels.
Thus, we also calculated the inclination and meridional inclination using only those pixels where $|B_r|$ exceeds  1$\sigma$ threshold above the rotational average.
We also introduce a 1\% minimum threshold for the overall data coverage, so that among 101 rotations of data that we have available there should be at least 364 data points in any latitudinal bin to be included in analysis.
Without this requirement the results at high latitudes would be dominated by single point outlier values, since there are typically very  few (if any) data points where $B_r$ exceeds the 1$\sigma$ threshold.
This limitation basically excludes all high-latitude observations, since $|B_r|$ very seldom exceeds the overall 1$\sigma$ threshold poleward of $\pm 35^\circ$ latitude. 

The red curves in Figure \ref{Fig6} shows the mean inclination and meridional inclination derived from all data points in each latitude bin (left panels of Figure \ref{Fig4}).
Black curves show the mean inclination and mean meridional inclination derived after excluding weak fields according to the method described above (the so-called strong field method).
Blue curves show the mean inclination and mean meridional inclination derived from the corresponding rotational means of the synoptic maps (corresponding to the middle panels of Figure \ref{Fig4}).

In agreement with Figure \ref{Fig4}, the overall mean inclination has a minimum around active regions, around latitudes of about +$19^\circ$ and  $-22^\circ$, increasing toward poles and equator. 
Meridional inclination shows the same smooth latitudinal trend   observed in  Figure \ref{Fig3}.
Meridional inclination follows an almost  linear trend, reaching a value of about $\pm 15^\circ$ at about $\pm 70^\circ$ latitude.

For strong fields (black curve), the mean inclination is systematically smaller than the mean inclination. 
The minimum is located at the equator and the maximum in the active region belt, but the latitudinal variation is quite weak, all values lying between 8$^\circ$ and 10$^\circ$.
The average meridional inclination indicates a roughly similar behavior for strong fields and all fields.
On average, from the equator up to about $\pm 35^\circ$ latitude, the strong fields are a few degrees more inclined toward the equator than all fields.
Accordingly, strong fields are systematically less inclined, but their meridional inclination is larger and more systematic.
Thus, the strong field inclination is typically oriented in the meridional direction.

Inclination of the field derived from rotational means (blue curve) is considerably larger than the all field mean inclinations  up to about $\pm 40^\circ$.
Mean field inclination has a maximum value of about  $30^\circ$ in active regions, but the latitudinal trend is weak in this latitude range. 
Mean field meridional inclination lies roughly between the all field strong field means up to about $\pm30^\circ$, but at higher latitudes it is smaller than all field meridional inclination. 

The result that mean inclination of strong fields is smaller than inclination derived from rotational means can be surprising, but can be understood as follows.
The two ends of an active region flux tube are only slightly tilted, but the latitudinal difference between the two poles is relatively small.
Active regions cover only a minor fraction of the solar surface and therefore do not have a significant effect on the longitudinal mean values.
However, the strongest fields dominate the longitudinal means of the magnetic field, as seen in Figure \ref{Fig2}.
Therefore, also total inclination and meridional inclination derived from longitudinal means are dominated by the most intense active regions (see Fig. \ref{Fig4}).

\section{Discussion}

The recently published synoptic maps of the photospheric vector field offer novel information about solar magnetic topology and connectivity.
The need for such data products was recognized  several decades ago, but regular daily synoptic observations of the full-disk vector magnetic field started only in 2010 by the SOLIS/VSM.

Based on the analysis of these new maps, we found that the radial component $B_r$ of the vector field is typically quite similar to $B_r^{PFSS}$ derived from  $B_{LOS}$ under the radial field assumption (see Figure \ref{Fig1}).
However,  vector $B_r$ is quite noisy at high latitudes.
This is due to a large contribution from transverse magnetic fields, which are noisier than the LOS field measurements.
Meridional vector field component $B_\theta$ resembles the potential field $B_\theta^{PFSS}$ in intense active regions, but in the decaying active regions and weak fields they tend to disagree. 
Similarly, polarities of the zonal vector field component $B_\phi$ and $B_\phi^{PFSS}$ agree in active regions, but disagree elsewhere.
In intense active regions, where field magnitude is typically larger than 100~G, the orientation of the field mainly reflects the orientation of the two ends of the active region. 
In intermediate fields, where the magnitude is typically tens of Gauss, the pattern of vector field suggests for the topology of a magnetic arcade. 
However, at least in some cases, the plane of the arcade field was inclined equatorward relative to local radial direction. 
Visual examination of the corresponding AIA images supported the presence of an inclined arcade field structure.

The supersynoptic map of the vector $B_r$  was found to have a closely similar structure as $B_r^{PFSS}$ (see Fig. \ref{Fig2}).
Longitudinal averages of the vector $B_\theta$ are weak and noisy outside the active region belts, but the strongest poleward surges of magnetic flux, especially the one in the south in 2014, are seen.

Zonal vector field component $B_\phi$ of the active regions has the orientation expected from the Hale rule.
The zonal $B_\phi$ component outside active regions is negative in the north and positive in the south in 2010-2015; thereafter, the  field is very weak in the north (but appears to be positive) and it is negative in the south. 
These results are quite unexpected since results based on $B_\phi$ derived from $B_{LOS}$ suggest that the zonal field of the next solar cycle appears at high latitudes soon after sunspot maximum \citep[e.g.,][]{Shrauner_1994,Ulrich_2005,Lo_2010}.
Our results partly agree with \cite{Pipin_2014}, who found negative $B_\phi$ in the south  in 1997 - 2009.
However, in 2010 our results suggest negative (positive) $B_\phi$ in the north (south) high latitudes, while \cite{Pipin_2014} show systematic positive $B_\phi$ in the north and negative in the south. 
According to Leighton's flux transport dynamo theory the zonal cycle starts at high latitudes right after the polar field reversal, when magnetic flux with new polarity radial field is transported equatorward from the pole in the convection zone \citep{Cameron_2017A}. 
While early studies based on $B_{LOS}$ largely agree with this model, our results based on vector magnetic field mainly disagree with it.
Further observations and studies are needed in order to understand the evolution of high latitude zonal fields and related dynamo processes.

We derived the inclination and meridional inclination of the vector magnetic field in each pixel of all synoptic maps. 
We studied the latitudinal and time evolution of the total inclination (I) and meridional inclination ($I_m$) of the vector magnetic field using three different methods: supersynoptic mean I and $I_m$, mean I and $I_m$ of supersynoptic magnetic field and strong field  method, where we selected only those pixels where $|B_r|$ exceeds 1$\sigma$ threshold above the average for each rotation.
It  should be noted that longitudinal averaging has a different effect on the three vector components, depending on their latitudinal position.
The radial field may largely cancel out in rotational averaging if the oppositely signed footpoints of an intense flux tube are located at the same latitude (non-tilted active region).
However, the zonal field may have  approximately the same orientation within an active region, which allows a large rotational mean of $B_\phi$.
Therefore, a rotational mean of vector components may depict a very inclined field due to the small $B_r$ and large $B_\phi$.
In the case of a tilted active region the two ends of a flux tube are at different latitudes and $B_r$ does not cancel out in rotational means.
In case of a unipolar structure (at high latitudes), averaging improves the signal-to-noise ratio, which makes the results more reliable.

SOLIS/VSM vector observations depict that the mean inclination is largest in weak fields, while the active regions show a roughly radial field structure.
However, inclination derived from supersynoptic magnetic field depicts largest values in active regions, and smallest inclination in unipolar fields (see Fig. \ref{Fig4}). 
This is somewhat surprising (taking into account the all field results), but understandable, since polar fields are known to be roughly radial and active region fields are obviously non-radial.
However, the relative size of the different effect of averaging in different vector components strongly depends on the way the data is treated, in particular on the averaging method, as discussed earlier in this paper.

The meridional inclination is found to be toward the equator in both hemispheres during the whole time interval studied (2010 - 2017). 
Average meridional inclination increases from zero at the equator toward the poles (see Figs. \ref{Fig4} and \ref{Fig6}).
The meridional inclination has a roughly similar latitudinal behavior even if we exclude weak fields.
Meridional inclination derived from longitudinal means of the vector field also show the same structure.
This systematic latitudinal pattern reflects a large-scale dipolar field, which becomes the dominant structure in the corona.

The total inclination and meridional inclination of the SOLIS/VSM vector magnetic field observations show systematic and persistent patterns of the large-scale photospheric magnetic field that cannot be observed using LOS magnetograph.
Figure \ref{Fig3} depicts that the distribution of inclination and meridional inclination have a clear and systematic maximum until about $\pm 50^\circ$  latitude.
In addition, Figure \ref{Fig6} shows that the meridional inclination patterns are systematic, regardless of the method used, until about $\pm 40^\circ$  latitude.

It is still unclear how the noise of the transverse component affects the vector magnetic field observations outside active regions. 
The transverse field does not have a sign and, therefore, the noise always increases the transverse component.
The expected pattern is that the noise of the transfer field makes the vector field less radial and more inclined in the direction away from the disk center. 
In this respect our finding of equatorward inclination should not be only a consequence of noise, since noise would prefer poleward inclination.
We note that our results are not directly comparable with the seemingly contradictory results of poleward inclination at the highest latitudes found in \cite{Ulrich_2013} since the high-latitude vector field observations have large uncertainties that are not understood in detail.
This will be  studied in greater detail in the future, when we expect more accurate uncertainty estimates of vector field data. 
The question of systematic high-latitude field inclination is extremely important since even a small systematic variation from radial direction would directly affect the radial flux estimated from LOS observations, and thus the coronal models and estimates of coronal open flux.

The magnitudes of $B_\phi^{PFSS}$ and $B_\theta^{PFSS}$ are larger than $B_\phi$ and $B_\theta$, while the $B_r^{PFSS}$ magnitude is quite close to $B_r$ (see Figs. \ref{Fig1} and \ref{Fig2}). 
This leads to a greater inclination of the PFSS field, which is largest in active regions and decreases poleward (see Fig \ref{Fig4}).
Inclination of the PFSS field does not significantly vary over the period studied. 
Meridional inclination of the PFSS field is typically equatorward from the equator up to about $\pm 20^\circ$ latitude and then poleward from $\pm 20^\circ$ to $\pm 50^\circ$.
From $\pm 50^\circ$ to the pole the PFSS field is typically inclined toward the equator, but there are periods, especially 2015 -- 2016 in the south, when field is inclined poleward even at the highest latitudes. 
These latitudinal patterns of the PFSS field reflect a structure where magnetic loops are closing between the northern and southern low latitudes (over the equator), but also from mid-latitudes to high latitudes within one hemisphere. 
However, eclipse observations \citep{Wiegelmann_2015} support instead the vector field observations and the idea of roughly radial magnetic fields in the photosphere.
This indicates that the limitations of the PFSS model (e.g., lack of currents and plasma) make the magnetic field close at too low altitudes and therefore lead to a more inclined field than observed. 

As already pointed out in the Introduction, the amount of meridional inclination is very important since  $B_r^{ps}$ is derived from  $B_{LOS}$ assuming that the field is radial.
Our current understanding of  photospheric and coronal magnetism (like the PFSS field) and space weather and space climate modeling relies on  $B_r^{ps}$ synoptic maps. 
The radial field hypothesis was introduced more that 40 years ago, and since that time has practically become a  self-evident truth.
This assumption actually disagrees with the properties of the PFSS model, where $B_\phi^{PFSS}$ and $B_\theta^{PFSS}$ are non-zero.
Our results in this paper show that equatorward meridional inclination makes $B_\theta$  contribute to $B_{LOS}$ increasingly at high latitudes, which makes the pseudo-radial field less noisy, but also slightly overestimated.
According to Figure \ref{Fig6}, the average meridional inclination is about 10$^\circ$ at $\pm 50^\circ$ latitude.
This would correspond to about a 20\% decrease in $B_r^{ps}$ relative to radial field assumption at $\pm 50^\circ$.

In addition to the systematic patterns of inclination and meridional inclination, vector magnetic field observations also show interesting variable features that relate to the evolution of photospheric magnetic field during SC 24. 
Figure \ref{Fig2} depicts the boundary between the flux transported from the decaying ARs ($B_r <0$ and $B_\theta > 0$) and the oppositely signed ($B_r > 0$ and $B_\theta < 0$) polar fields in the southern hemisphere in 2012- 2014. 
The middle panels of Figure \ref{Fig4} show that the magnetic field in this boundary region is more horizontal than the surrounding fluxes of opposite polarity.
The location of this boundary corresponds to the filament channel of polar crown filaments that are associated with polar field reversals \citep{Gopalswamy_2016}.

One of the obvious issues with vector field observations is the vantage point effect (due to the varying $b_0$-angle), which distorts vector $B_r$ observations already around  60$^\circ$ latitude, while $B_{LOS}$ observations do not suffer from this problem until about 80$^\circ$.
The main problem in measuring $B_r^{ps}$ at high latitudes is the decreasing projection of the the radial field to the LOS direction toward the pole.
On the other hand, the vector field $B_r$ measured at high latitudes mainly reflects the transverse component of the magnetic field, which is  noisier than  the longitudinal field. 
It seems that the latter problem is more severe since the vantage point effect is considerably more significant in the vector $B_r$.
Moreover, the vantage point effect can be seen to affect the inclination even at low latitudes and meridional inclination even at the equator (see Fig. \ref{Fig4}).
It seems that coronal models in the near future will still have to rely on $B_r^{ps}$, or possibly a combination of $B_r^{ps}$ (at high latitudes / weak fields) and the vector $B_r$ (at low latitudes / active regions).

\section{Conclusions}

The vector field synoptic maps of photospheric magnetic field from SOLIS/VSM have been available since 2010 and provide new  insights into the evolution and structure of solar magnetic fields.

Our analysis of this data shows that the photospheric magnetic field is in general fairly non-radial. 
The effect of this on current (space weather) modeling needs to be further investigated as the pseudo-radial fields employed by many modern models (e.g., PFSS, WSA-Enlil) are derived from LOS magnetograms under the assumption that the magnetic field  in the photosphere is radial.
Comparing vector and PFSS fields in the photosphere shows agreement only for the large-scale radial field. 
Zonal and meridional vector field components $B_\theta$ and $B_\phi$ agree with PFSS field components $B_\theta^{PFSS}$ and $B_\phi^{PFSS}$ only in newly emerged strong active regions. 
PFSS field components $B_\theta^{PFSS}$ and $B_\phi^{PFSS}$ show considerably larger magnitude than $B_\theta$ and $B_\phi$, which is also seen as a larger inclination of the PFSS field.
Our results show that the photospheric magnetic field does not exhibit strong super-radial expansion and is more radial than the potential field. 
This difference can be most likely explained in terms of plasma flows and currents, which are not included in the PFSS field.
Moreover, B$_r^{ps}$ , which is used as a boundary condition in the PFSS model, is based on a constant (unity) filling factor used in LOS measurements. 

Our results show that the inclination of the photospheric magnetic field is smallest in active regions and increases toward the poles and equator. 
When inclination is derived using only strong fields or longitudinally averaged vector fields, the latitudinal variation is weak and inclination is largest in active regions. 
This difference is  partly a consequence of the higher noise of weak fields $B_\theta$ and $B_\phi$, which intensifies the horizontal field and makes it more inclined. 
When weak fields are removed or smoothed by averaging, the inclination is slightly larger in strong fields. 
Meridional inclination patterns are systematic and do not have a significant solar cycle evolution. 
The fields are, on an average, inclined toward the equator and are similar for strong fields or when using longitudinally averaged vector fields.
However, the meridional inclination derived from longitudinal averages depict more variability and  is smallest in the poleward surges of unipolar magnetic field.
This latitudinal pattern of the meridional inclination of the magnetic field reflects the dipolar structure of the solar magnetic field, and should be taken into account when deriving $B_r^{ps}$ component from $B_{LOS}$.

Current measurements of vector magnetic fields exhibit several deficiencies that need to be  addressed in future observations. 
Specifically, we need to develop a better understanding of the effect of noise on the derived intensity and orientation of the vector magnetic field.
A successful implementation of vector fields in modeling also requires addressing the pole filling issue.
Polar field vector observations are very noisy and suffer from the vantage point effect considerably more than the LOS field observations.
This could be improved by vector observations of high-latitude fields with better signal-to-noise ratio, or by combining the B$_r$ component from vector measurements at low to middle latitudes with B$_r^{ps}$ observations for polar areas.
\\
\\
\\

\begin{acknowledgements}

We acknowledge the financial support by the Academy of Finland to the ReSoLVE Centre of Excellence (project no. 307411).
Data were acquired by SOLIS/VSM instruments operated by NISP/NSO/AURA/NSF.
SOLIS: ftp://solis.nso.edu/integral/kbv9g
AIA data used in this study courtesy of NASA/SDO and the AIA science team.
A.A.P. acknowledges NASA grants NNH14AX89I and NNX15AN43G for the development and production of synoptic vector maps.
The authors are members of international team on Reconstructing Solar and Heliospheric Magnetic Field Evolution Over the Past Century supported by the International Space Science Institute (ISSI), Bern, Switzerland.
The authors thank the referee, Roger Ulrich, for the constructive comments that improved this article.
\\
\end{acknowledgements}

%

\end{document}